\documentclass[pdftex,twocolumn,epjc3]{svjour3}          

\RequirePackage[T1]{fontenc}

\smartqed  

\RequirePackage{graphicx}
\RequirePackage{mathptmx}      
\RequirePackage{flushend}
\RequirePackage[numbers,sort&compress]{natbib}
\RequirePackage[colorlinks,citecolor=blue,urlcolor=blue,linkcolor=blue]{hyperref}

\journalname{Eur. Phys. J. C}
\usepackage{graphicx}
\usepackage{amssymb}
\usepackage{amsmath}
\usepackage{color}
\usepackage{hyperref}
\usepackage{tabu}

\def\barray{\begin{array}}
	\def\earray{\end{array}}
\def\be{\begin{equation}}
	\def\ee{\end{equation}}
\def\ben{\begin{equation} \nonumber}
	\def\een{\end{equation}}
\def\ban{\begin{eqnarray*}}
	\def\ean{\end{eqnarray*}}
\def\ba{\begin{eqnarray}}
	\def\ea{\end{eqnarray}}

\def\({\left(}
\def\){\right)}

\graphicspath{{./fig/}}

\begin{document}

\title{Dark energy reconstruction based on the Pad\'{e} approximation; an expansion around the $\Lambda$CDM}
	\author{Ahmad  mehrabi\thanksref{e1,ad1},
		Spyros Basilakos\thanksref{e2,ad2,ad3}
	}
		\thankstext{e1}{e-mail: Mehrabi@basu.ac.ir}
	\thankstext{e2}{e-mail:svasil@academyofathens.gr}

	\institute{Department of Physics, Bu-Ali Sina University, Hamedan
		65178, 016016, Iran \email{e-mail: Mehrabi@basu.ac.ir} \label{ad1}
		\and
		Academy of Athens, Research Center for Astronomy \& Applied
		Mathematics, Soranou Efessiou 4, 11-527, Athens, Greece \label{ad2}
		\and
		National Observatory of Athens, Lofos Nymfon, 
		11852, Athens, Greece \label{ad3}
	}
	
	\date{Received: date / Accepted: date}

	\maketitle

\begin{abstract}
We study the dynamical properties of dark energy 
based on a large family of Pad\'{e} parameterizations for which 
the dark energy density evolves as a ratio between 
two polynomials in the scale factor of the universe.
Using the latest cosmological data
we perform a standard likelihood analysis in order to
place constraints on the main cosmological parameters 
of different Pad\'{e} models. 
We find that the basic cosmological 
parameters, namely $(\Omega_{m0},h,\sigma_{8})$ are practically the same
for all Pad\'{e} parametrizations explored here. 
Concerning the free parameters which are related to dark energy 
we show that the best fit values indicate that the equation of state 
parameter at the present time is in the phantom regime ($w<-1$), however
we can not exclude the possibility of $w>-1$ at $1\sigma$ level. 
Finally, for the current family of Pad\'{e} parametrizations 
we test their ability, via AIC, BIC and Jeffreys' scale, to deviate 
from $\Lambda$CDM cosmology.
Among the current Pad\'{e} parametrizations, the model which 
contains two dark energy parameters is the one 
for which a small but non-zero deviation from
$\Lambda$CDM cosmology is slightly allowed by AIC test.
Moreover, based on Jeffreys' scale  
we show that a deviation from $\Lambda$CDM cosmology
is also allowed and thus 
the possibility of having a dynamical dark energy  
in the form of Pad\'{e} parametrization 
cannot be excluded.

\end{abstract}

\section{Introduction}\label{intro}
It is well known that 
the concept of dark energy (DE) was introduced in order to describe the
accelerated expansion of the Universe. Therefore, understanding 
the nature of DE is considered one of the most difficult and 
fundamental problems in cosmology.
The introduction of a cosmological constant, $\Lambda$ ($\rho_{\Lambda}=$const.), 
is perhaps the simplest form of DE which can be considered \citep{Peebles2003}. 
The outcome of this consideration is the 
concordance $\Lambda$CDM model, for which $\Lambda$ 
constant coexists with cold dark matter (CDM) and baryonic matter. 
In general, this model 
provides a good description of the observed 
Universe, since it is consistent with the cosmological data, namely 
Cosmic Microwave Background (CMB) \citep{Komatsu2009,Jarosik:2010iu,Komatsu2011,Ade:2015yua}, Baryon Acoustic Oscillation (BAO) \citep{Tegmark:2003ud,Cole:2005sx,Eisenstein:2005su,Percival2010,Blake:2011rj,Reid:2012sw} and Supernovae Type-Ia (SnIa)\citep{Riess1998,Perlmutter1999,Kowalski2008,Betoule:2014frx}.
Despite the latter achievement $\Lambda$CDM suffers from 
the cosmological constant and the coincidence problems \citep{Weinberg1989,Sahni:1999gb,Carroll2001,Padmanabhan2003,Copeland:2006wr}. 
A third possible problem is related with the fact that 
the determination of the
Hubble constant and the mass variance at $8h^{-1}$Mpc
have indicated a tension between
the values provided by the analysis of Planck data 
and the results obtained by 
the late time observational data 
\citep{Abdalla:2014cla,Riess:2016jrr,Hildebrandt:2016iqg}. 

An alternative avenue to overcoming the above problems is to 
introduce a dynamical DE, wherein the density of DE is allowed to 
evolve with cosmic time \citep{Caldwell:1997ii,Erickson:2001bq,Armendariz2001,Caldwell2002,Padmanabhan2002,Elizalde:2004mq,Zhao:2017cud}. 
The first choice is to consider a DE fluid where the equation of state 
parameter varies with redshift, $w(z)$. 
Usually, in these kind of studies the EoS parameter can be written either 
as a first-order Taylor expansion 
around $a(z)=1$ \citep{2001IJMPD..10..213C,Linder2003} or as a Pad\'{e} parametrization 
\citep{Adachi:2011vu,Gruber:2013wua,Wei:2013jya,2017ApJ...843...65R}, where the corresponding free parameters are fitted by the cosmological data
\citep{Riess2004,Seljak:2004xh,Bassett:2007aw,Dent:2008ek,Frampton:2011rq,Feng:2012gf,Mehrabi:2014ema,Mehrabi:2015hva}.
Notice that for $w>-1$ we are
in the quintessence regime \citep{ArmendarizPicon:2000dh,Copeland:2006wr}, 
namely the corresponding scalar field has a canonical 
Lagrangian form. In the case of $w<-1$ we are in the 
phantom region where the Lagrangian of the scalar field 
has a non-canonical form (K-essence) \citep{ArmendarizPicon:1999rj,ArmendarizPicon:2000dh,Chiba:1999ka,Chiba:2009nh,Amendola:2010}. 
On the other hand, it is possible to reconstruct a DE model directly 
from observations. This approach provides an
excellent platform to study DE and indeed one may find 
several attempts in the literature.
Specifically, one may use parametric criteria toward 
reconstructing directly the evolution 
of DE density $\rho_{\rm de}(z)$ \citep{Wang:2001ht,Maor:2001ku,Wang:2004ru}  
and the potential of the scalar field \citep{Mamon:2016wow}.  
\footnote{For non-parametric criteria we refer the reader the following papers 
\citep{Holsclaw:2011wi,Sahlen:2005zw,Sahlen:2006dn,Crittenden:2011aa}.}
Comparing the two methods, namely $w(z)$ and $\rho_{\rm de}(z)$ 
for the same observational data-sets, it has been found that 
the latter method provides tighter constrains on the free parameters 
than the former \citep{Wang:2001ht,Maor:2001ku,Wang:2004ru}.  

In this work we have decided to reconstruct the evolution of the 
DE density, using the well known Pad\'{e} approximation for which 
an unknown function \citep{PADE.book,2012PThPh.127..145A} 
is well approximated by the ratio of  
two polynomials. In contrast to the case where the Pad\'{e} approximation use to describe the DE EoS, our approach provides an interesting parameterization which can be regarded as an expansion around the $\Lambda$CDM.  
In section (\ref{sec:formalism}) we 
introduce the concept of Pad\'{e} approximation in DE cosmologies. 
In section (\ref{sec:bayes}) we briefly discuss the main features of 
the Bayesian analysis used in this work as well as we briefly 
present the observational data. 
In section  (\ref{sec:res}) we discuss the main results of our work, namely 
we provide the observational constraints on the fitted model parameters 
and we test whether a dynamical DE is allowed by the current data.
Finally, in section (\ref{sec:con}) we provide our conclusions.

\section{Reconstruction of dark energy using Pad\'{e} approximation }\label{sec:formalism}
\subsection{background evolution}
Considering an isotropic and homogeneous Universe, driven 
by radiation, non-relativistic matter and dark energy 
with equation of state, 
$P_{\rm Q}=w(a)\rho_{\rm Q}<0$, the first Friedman equation is given by:
\begin{equation}
H^{2}=\left( \frac{\dot{a}}{a} \right)^{2}=
\frac{8\pi G}{3}(\rho_{r}+\rho_{m}+\rho_{\rm Q})-\frac{k}{a^{2}}
\end{equation}
with $k=-1, 0$ or 1 for
open, flat and closed universe respectively. 
For the rest of the analysis we have set $k=0$.
Notice that $a(t)$ is the scale 
factor, $\rho_{r}=\rho_{r0} a^{-4}$ is the radiation density, 
$\rho_{m}=\rho_{\rm m0} a^{-3}$ is the 
matter density and 
$\rho_{\rm Q}=\rho_{\rm Q0} X(a)$ is the dark 
energy density, with:
\begin{equation}
X(a)={\rm exp}\left[ 3\int_{a}^{1} \left(
  \frac{1+w(u)}{u}\right) {\rm d}u \right] \;\;.
\end{equation}
or
 \begin{equation}\label{eq:eos}
 w(a)=-1 + \frac{1}{3}a\frac{X^{\prime}}{X},\;
 \end{equation} 
where the prime denote derivative with respect to the scale factor. 
Combining the above equation we easily obtain the normalized 
Hubble parameter $E(a)=H(a)/H_{0}$ 
\begin{equation}
E(a)=\left[ \Omega_{r0}a^{-4}+\Omega_{m0}a^{-3}+
\Omega_{\rm Q0 }X(a)\right]^{1/2} \;\;\;,
\end{equation}
where $\Omega_{r0}= 8\pi G \rho_{r0}/3H_{0}^{2}$ (radiation density parameter), 
$\Omega_{m0}= 8\pi G \rho_{m0}/3H_{0}^{2}$, 
(matter density parameter), 
$\Omega_{\rm Q0}= 8\pi G \rho_{\rm Q0}/3H_{0}^{2}$ 
(DE density parameter) at the present time with
$\Omega_{r0}+\Omega_{m0}+\Omega_{\rm Q0}=1$.
Since the physics of DE is still an open issue 
the function $X(a)$ encodes our ignorance concerning 
the underlying mechanism powering the late time
cosmic acceleration. Of course for $X(a)=1$ we recover the 
concordance $\Lambda$CDM model, namely $w=-1$.

In order to investigate possible deviations from the concordance 
$\Lambda$ cosmology, we consider an expansion of the 
function $X(a)$ using the so called Pad\'{e} approximation. 
In general, for an arbitrary function $f(x)$ the 
Pad\'{e} approximation of order $(n,m)$ 
is given by \citep{PADE.book,2012PThPh.127..145A}
      \begin{equation}\label{eq:Pade}
      f(x)=\frac{b_0 + b_1x+b_2x^2+b_3x^3+...+b_nx^n}{c_0 + c_1x+c_2x^2+c_3x^3+...+c_mx^m},\;
      \end{equation} 
where the exponents are positive and the corresponding
coefficients $(b_i,c_i)$ are constants. 
Obviously, in the case of $c_i=0$ ($i>0$)
the above expansion reduces to the usual Taylor expansion. 
One of the main advantages of such an approximation 
is that by considering the same order 
for $m=n$ the Pad\'{e} approximation 
tends to finite values at both $x\rightarrow \infty$ 
and $x\rightarrow 0$ cases.  
   
Based on the above formulation the unknown $X(a)$ function
is approximated by  
         \begin{equation}\label{eq:x_a-Pade}
         X(a)=\frac{b_0 + b_1(1-a)+b_2(1-a)^2+...+b_n(1-a)^n}{b_0 + c_1(1-a)+c_2(1-a)^2+...+c_m(1-a)^m},\;
         \end{equation} 
where we have set $x=1-a$ and $c_0=b_0$ as a result of $X(a=1)=1$. 
In order to simplify further the calculation, 
we may cancel $b_0$ 
from both numerator and denominator and rename the corresponding 
constants. Therefore, we have  
   \begin{equation}\label{eq:x_a-Pade2}
      X(a)=\frac{1+ b_1(1-a)+b_2(1-a)^2+...+b_n(1-a)^n}{1 + c_1(1-a)+c_2(1-a)^2+...+c_m(1-a)^m}.\;
    \end{equation} 

Using Eq.(\ref{eq:x_a-Pade2}) and
differentiating $X(a)$ with respect to scale factor 
the EoS parameter (\ref{eq:eos}) takes the following form
   \begin{equation}\label{Gen}
w=-1-\frac{1}{3}a\left[ \frac{\sum_{i=1}^{n}ib_{i}(1-a)^{i-1}}
{1+\sum_{i=1}^{n}b_{i}(1-a)^{i}}-
\frac{\sum_{i=1}^{m} ic_{i}(1-a)^{i-1}}
{1+\sum_{i=1}^{m}c_{i}(1-a)^{i}}\right] \;.
    \end{equation} 
Here we use the Pad\'{e} approximation to model the energy density of DE rather than the EoS \citep{2017ApJ...843...65R}. Our approach provides a new and interesting  parameterization which can be regarded as an expansion around the $\Lambda$CDM.
        
Inserting $a=1$ into the latter equation we obtain the EoS parameter
at the present time, namely
 \begin{equation}\label{eq:eos-per}
      w_{0}\equiv w(a=1)=-1 + \frac{1}{3}(c_1-b_1)\; .
 \end{equation}
Interstingtly, in the case of $b_1>c_1$ 
the current value of $w_{0}$ 
can cross the phantom line $w_{0}<-1$, while for $b_1<c_1$  
it remains in the quintessence regime $w_{0}>-1$. 
Moreover, 
for $b_{i}=c_{i}=0$ 
we recover the $\Lambda$CDM model, while 
for $b_{i+1}<b_i$ and $c_{i+1}<c_i$ the current family of Pad\'{e} models 
can be seen as an expansion around the $\Lambda$CDM where the '
extra terms indicate a dynamical DE. 
Unlike most DE parametrizations (CPL and the like), here it is easy to show 
that the EoS
parameter avoids the divergence in the far future, hence it is a 
well-behaved function in the range of $a\in (0,+\infty)$.
Keeping the leading terms $(b_{1},c_{1})$ in Eq.(\ref{Gen}) 
we arrive at 
 \begin{equation}\label{eq:wa}
 	w(a) = -1 +a\frac{1}{3}\frac{c_1-b_1}{(1+c_1(1-a))(1+b_1(1-a))}.
 \end{equation}

\begin{figure}
 	\centering
 		\includegraphics[width=.48\textwidth]{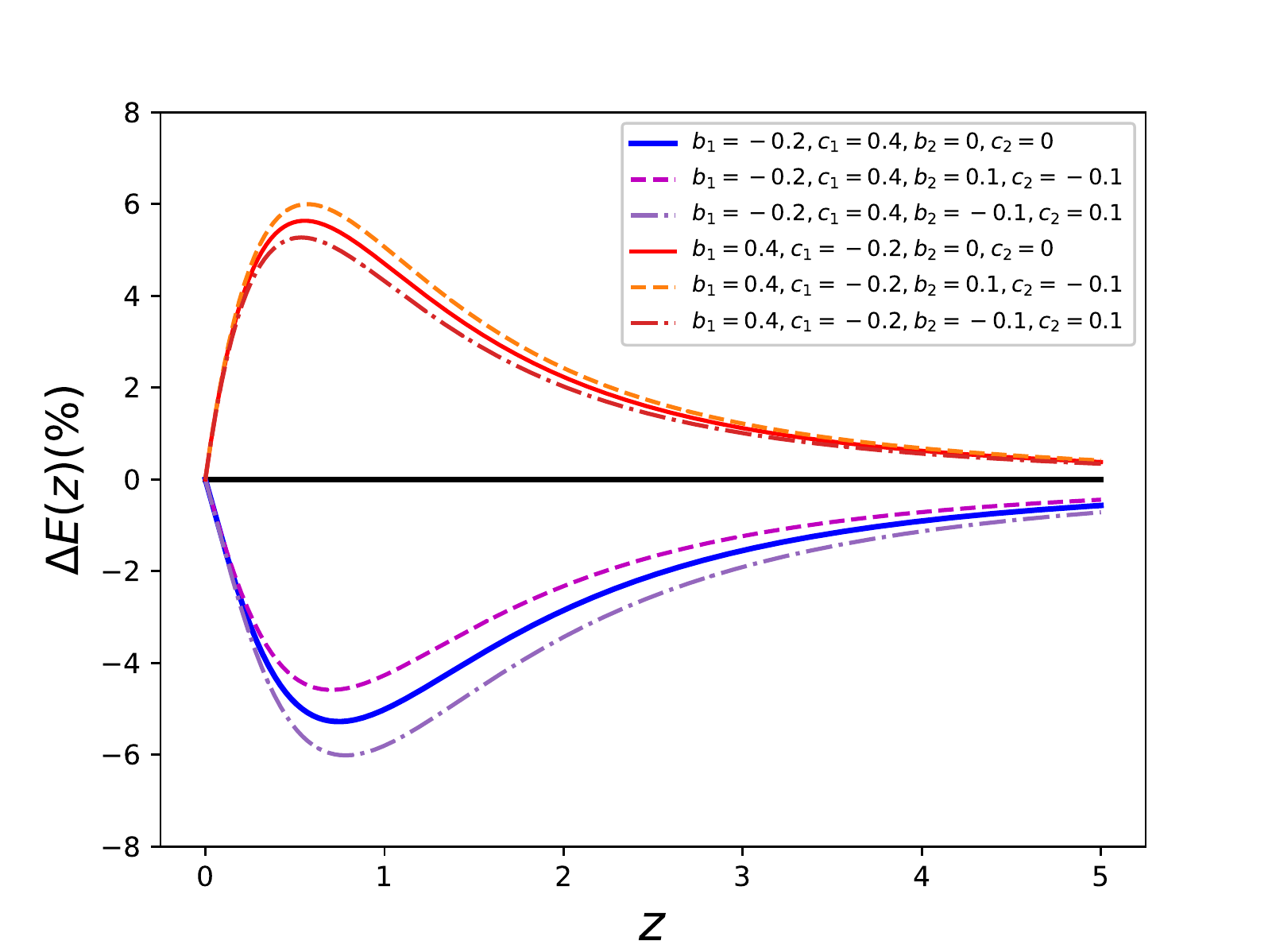}
\caption{The relative difference of the Hubble parameter 
using some typical values of the model parameters. 
$\Delta E$ becomes positive (negative) when $b_1>c_1$ $(b_1<c_1)$.}
\label{fig:deltaE}
 \end{figure}

To visualize the differences of various Pad\'{e} 
models with respect to the expectations of the usual 
$\Lambda$CDM model we plot 
in Fig.(\ref{fig:deltaE}) the corresponding relative differences 
 \begin{equation}\label{eq:rel-E}
    \Delta E(\%)=\frac{E(z)-E_{\Lambda}(z)}{E_{\Lambda}(z)}\times 100.
 \end{equation} 
For simplicity the matter density parameter is fixed to 
$\Omega_{m0}=0.3$. Using Eq.(\ref{eq:rel-E}) and
in the case of $b_{1}>c_{1}$ we have $\Delta E>0$ and  
the present value of the EoS parameter is in the phantom regime 
($w<-1$). Notice, that the opposite holds for $b_{1}<c_{1}$. 
Moreover, prior to $z\sim 1$ we find $\pm 6\%$ Hubble 
function differences, while 
$\Delta E$ tends to zero at high redshifts.
Lastly, we would like to illustrate how extra terms of $X(a)$ 
affect the Hubble parameter. As an example, we introduce the quantity 
$(1-a)^{2}$ in Eq.(\ref{eq:x_a-Pade}), where regarding the 
corresponding constants we have set them either to $(b_{2},c_{2})=(0.1,-0.1)$ 
or $(b_{2},c_{2})=(-0.1,0.1)$. 
In Fig.(\ref{fig:deltaE})
we present for the above set of $(b_{2},c_{2})$ parameters 
the evolution of $\Delta E$. 
It is obvious from the figure 
that the extra term $(1-a)^{2}$ in the 
function $X(a)$ does not really affect the cosmic expansion.

Consequently, our model can be considered as an 
expansion around the $\Lambda$CDM in the sense that adding extra 
terms change slightly the Hubble parameter 
   
\subsection{Growth of perturbations}
It is well known that DE not only affects the expansion rate of the universe 
but also it has an impact on the growth rate of matter perturbations. 
In order to realize how different forms of Pad\'{e} 
parametrizations affect the growth rate of fluctuations, we solve the perturbed 
equations and compare the solution with that of $\Lambda$CDM model. 
Assuming a homogeneous DE fluid, the evolution of matter 
perturbations in the linear regime are given by \citep{Abramo:2008ip}
  \begin{equation}\label{eq:per1}
     \dot{\delta_m}+\frac{\theta_m}{a}=0,\;
    \end{equation}  
   \begin{equation}\label{eq:per2}
   \dot{\theta_m}+H\theta_m - \frac{k^2\phi}{a}=0,\;
   \end{equation}  
where the dot denote derivative with respect to cosmic time, $\delta_m$ 
is the overdensity contrast and 
and $\theta_m$ is the velocity divergence.   
Combining the above set of equations with the Poisson equation 
  \begin{equation}\label{eq:pos}
  -\frac{k^2\phi}{a^2}=\frac{3}{2}H^2\Omega_m\delta_m,\;
  \end{equation}  
we find after some calculations   
   \begin{equation}\label{eq:per12}
   \delta_m^{\prime\prime}+\frac{1}{a}(3+\frac{\dot{H}}{H^2})\delta_m^{\prime}=\frac{3}{2}\Omega_m\delta_m,\;
   \end{equation} 
where $\Omega_m=\frac{\Omega_{m0} a^{-3}}{E^2(a)}$. 
Notice that the latter differential 
equation is written in terms of the scale factor, hence 
   \begin{equation}\label{eq:hdot}
     \frac{\dot{H}}{H^2}=\frac{1}{2}a\frac{(E^2)^{\prime}}{E^2}.\;
   \end{equation}
Therefore, for those cosmological models which 
are inside GR the linear matter 
perturbations are only affected 
by $E(a)$, while in the case of extended gravity models 
we need to modify the Poisson equation.

An important quantity toward testing 
the performance of the DE models at the perturbation level is 
$f\sigma_8(a)$, where $f=a\frac{\delta_m^{\prime}}{\delta_m}$ is the  
growth rate of clustering and 
$\sigma_8(a)$ is the mass variance inside a sphere of radius 
8$h^{-1}$Mpc. The mass variance is written as 
$\sigma_8(a)=\sigma_8\frac{\delta_m(a)}{\delta_m(a=1)}$, where 
$\sigma_{8}\equiv \sigma_8(a=1)$ is the corresponding value at the present time.
Notice, that in our work we treat  
$\sigma_8$ as a free parameter and thus it will be 
constrained by the available growth data. 

\begin{figure}
 	\centering
 	\includegraphics[width=.48\textwidth]{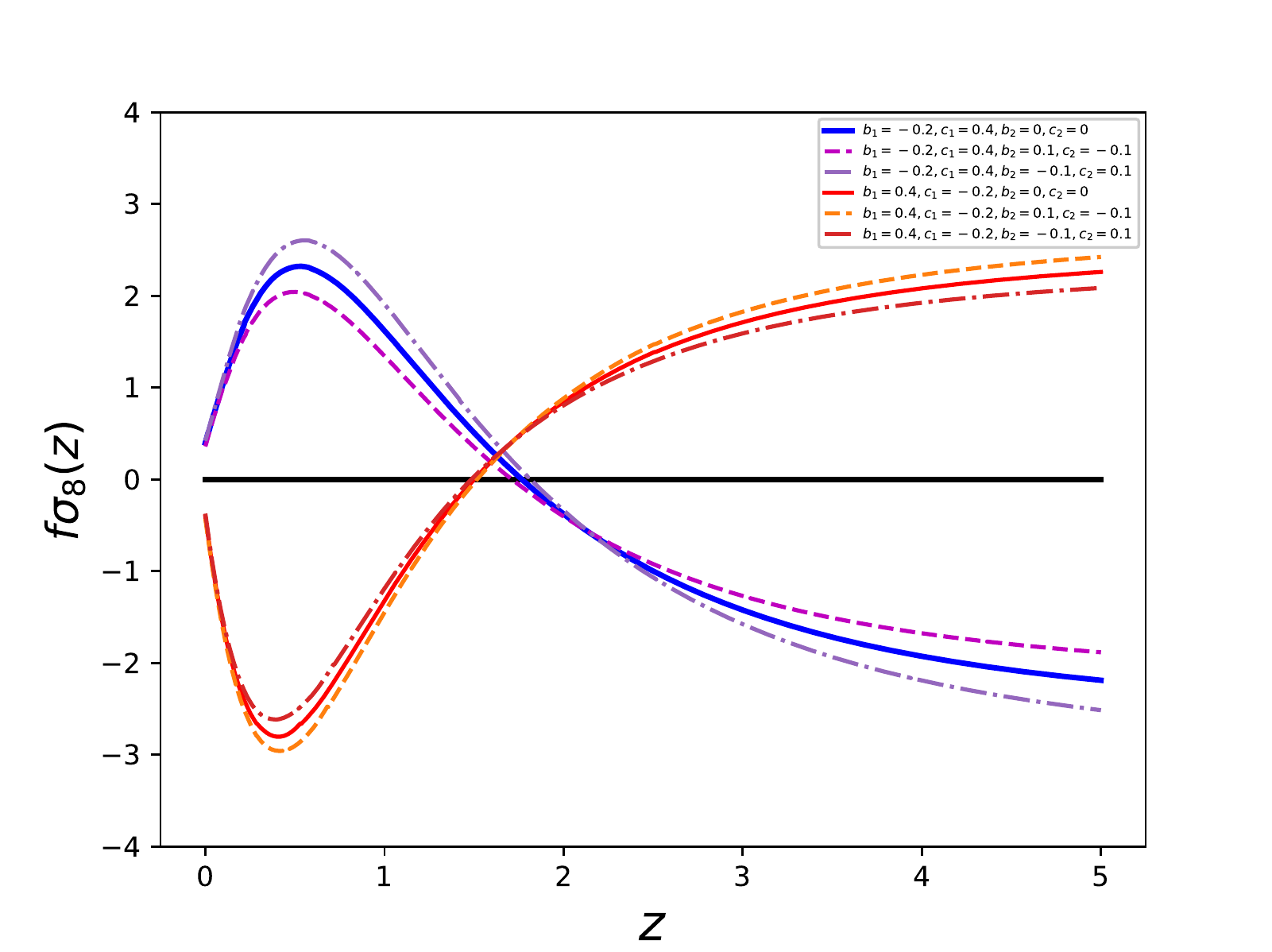}
 	\caption{The relative difference of growth rate 
as a function of redshift. The free parameters are 
the same with those of Fig.(\ref{fig:deltaE})}
 	\label{fig:fs8}
 \end{figure}

In order to understand the differences between $\Lambda$CDM and Pad\'{e} models 
at the perturbation level
in Fig.(\ref{fig:fs8}) we plot the relative fractional difference, namely
   \begin{equation}\label{eq:rel-fs8}
   \Delta f\sigma_8(\%)=\frac{f\sigma_8-(f\sigma_8)_{\Lambda}}{(f\sigma_8)_{\Lambda}}\times 100.\;
   \end{equation} 

For better comparison, the free parameters used  
in this figure are the same with those of 
Fig.(\ref{fig:deltaE}), where we have set  
$\Omega_{m0}=0.3$, $h=0.7$ and $\sigma_8=0.8$. 
Overall, for phantom Pad\'{e} cosmologies ($b_1>c_{1}$, see red line) we find that 
the expected differences are small at low redshifts, but they become 
larger for $z\simeq 0.5$, reaching variations of up to
$\sim -3\%$, while they turn to positive at high red-shifts. 
Notice that the opposite behavior holds in the case of 
quintessence Pad\'{e} cosmologies ($b_1<c_{1}$, see blue line).

\section{Bayesian evidence and data processing}\label{sec:bayes}
Using Bayes' theorem, it is possible to find the 
probability of a model in the light of 
given observational data. 
Given a data set (D) 
the probability of having a model (M) is 
    \begin{equation}\label{eq:baye-theo}
    P(M|D)=\frac{P(M|D)P(M)}{P(D)}.\;
    \end{equation}    
The posterior probability of the free 
parameters ($\theta$) of the model is given by
     \begin{equation}\label{eq:post-like}
     P(\theta|D,M)=\frac{P(D|\theta,M)P(\theta|M)}{P(D|M)},\;
     \end{equation} 
where $P(D|\theta,M)$ is the likelihood function 
of the model with its parameters 
and $P(\theta|M)$ is the prior information on the free parameters. 
For parameter estimations, we only need the 
likelihood function and the prior, hence the denominator 
which is a normalization constant has no impact 
on the value of free parameters. 
Practically, the denominator is the integral of 
the likelihood and prior product over 
the parameter space
  \begin{equation}\label{eq:evidence}
     {\cal{E}} = P(D|M)=\int d\theta P(D|\theta,M)P(\theta|M),\;
  \end{equation}       
The latter quantity has been widely used in the literature
\citep{Trotta:2008qt,Martin:2010hh,Saini:2003wq,Lonappan:2017lzt}, toward
selecting the best model from a given family of models.

With the aid of Pad\'{e} parametrization which  
can be seen as an expansion around $\Lambda$CDM ($w=-1$)
our aim is to check whether 
the current observational data prefer a dynamical DE.
First we consider a large body of Pad\'{e} parametrizations 
(\ref{eq:x_a-Pade2}) and then we test the statistical performance 
of each Pad\'{e} model against the data. 

Now, let us briefly present the observational data 
that we utilize in our analysis

 \begin{itemize}
 	\item We use the JLA SnIa data of (full likelihood version) \citep{Betoule:2014frx}. 
 	\item The Baryon Acoustic Oscillations (BAO) data  
from 6dF \citep{Beutler:2012px}, SDSS \citep{Anderson:2012sa} and WiggleZ \citep{Blake:2011en} surveys. Notice, that 
details of concerning the data processing and likelihoods can be found in \citep{Hinshaw:2012aka,Mehrabi:2015hva}.

 	\item The Hubble parameter measurements as a function of redshift. 
We utilize the $H(z)$ data set as provided by \citep{Farooq:2016zwm}.

 	\item The CMB shift parameters as measured by the Planck team \citep{Ade:2015rim}. 
Notice that we use the covariance matrix which is introduced in Table 4 of \citep{Ade:2015rim}.   

 	\item The Hubble constant from \citep{Riess:2016jrr}.
 	 
 	\item For the growth rate data, in addition to "Gold" growth dataset $f\sigma_8(z)$ provided by \citep{Nesseris:2017vor}, we also use 5 new data points as collected by \citep{Kazantzidis:2018rnb}. These new data points provide growth rate at relatively higher redshifts and there is no overlap between these data and the gold sample. 
These new data points and their references 
are presented in Table (\ref{tab:fs8}).
 \end{itemize}  

\begin{table}
	\centering
	\begin{tabular}{c|c|c}
		\hline
		\hline
		$z$     & $f\sigma_8$    & Survey \& References  \\ \hline
		$0.64$  & $0.486\pm0.070$ & BOSS DR12 \citep{Wang:2017wia}\\ \hline
		$1.52$  & $0.42\pm0.076$ & SDSS-IV \citep{Gil-Marin:2018cgo}\\ \hline
        $0.978$  & $0.379\pm0.176$ & SDSS-IV \citep{Zhao:2018jxv}\\ \hline
        $1.23$  & $0.385\pm0.099$ & SDSS-IV \citep{Zhao:2018jxv}\\ \hline
        $1.94$  & $0.364\pm0.106$ & SDSS-IV \citep{Zhao:2018jxv}\\ \hline		
	\end{tabular}
	\caption{New $f\sigma_8$ data points which we use along 
with the Gold sample.}	\label{tab:fs8}
\end{table}
 
Concerning the estimation of the sound horizon, needed when 
we compute the CMB and BAO likelihoods,
we follow the procedure of \citep{Wang:2013mha}.
 Using the aforementioned data sets, we first perform a MCMC 
analysis to find the best value of parameters as well as their 
uncertainties and then we quantify the statistical ability of
each model to fit the observational data.  
To do this we use the MULTINEST 
sampling algorithm \citep{Feroz:2008xx} and the python 
implementation \emph{pymultinest} \citep{Buchner:2014nha}. 
The latter technique was initially proposed in order to select 
the best model of AGN X-ray spectra via a Bayesian approach.

\section{Results and discussion}\label{sec:res}
As we have already mentioned 
nowadays, testing the evolution of the DE EoS parameter 
is considered as one of the most fundamental problems in cosmology.
We attempt to check such a possibility in the context of Pad\'{e} parametrizations.
Specifically, the family of Pad\'{e} 
models as well as the corresponding free parameters used here are shown in 
Tables (\ref{tab:models}) and (\ref{tab:prior}) respectively. Since, unlike parameter estimation, the evidence strongly depends on the prior, we consider two different priors to show how the evidence changes due to different prior ranges. Here we select flat priors which are often a standard choice. The upper panel in table(\ref{tab:prior}) shows a narrow range of priors while the lower panel presents a more wider prior. The priors on the cosmological parameters ($\Omega_{\rm dm},\Omega_{\rm ba},h,\sigma_8$) are physically reasonable due to our understanding from observational data including SN Ia, CMB, BAO and growth rate of large scale structures. On the other hand, the range of priors in ($\alpha,\beta, M ,\Delta M$) come from analysis of SN Ia data in \cite{Betoule:2014frx,Wang:2016bba}. In contrast, we have no prior information regarding our free parameters in the Pad\'{e} expansion ($b_1,c_1,b_2,c_2,b_3,c_3,b_4,c_4$). So, we select priors on these parameters from the intuition that they should construct an expansion around the $\Lambda$CDM model. 
In this sense, we consider smaller prior ranges for parameters which are higher order in the expansion. According to \cite{Trotta:2008qt}, one possibility in such a case is to consider a prior which maximizes the probability of the new model, given the data. In this case, if the evidence is not significantly larger than the simpler model, then we can say the data does not support additional parameters. In addition to these two prior ranges, we have examined other prior ranges and our results did not change significantly.
     		   
In table (\ref{tab:prior}), $\alpha,\beta$ and $M$ are nuisance parameters which are used to model the empirical distance modulus of each SN and $\Delta M$ is used to correct for dependency of the absolute magnitude in rest-frame B band, to the host stellar mass. For more information regarding definition of these parameters refer to \cite{Betoule:2014frx}. 
Therefore, using the cosmological data we place constraints 
on the model parameters but also we provide a visual
way to discriminate cosmological models.

\begin{table}
	\begin{center}
		\begin{tabular}{ |c|c| }
			\hline
			\hline 
			$M_1$ & $\Lambda$CDM \\ 
			\hline
			$M_2$ & $b_1,c_1\neq 0 $ and all others equal to zero \\ 
			\hline
			$M_3$ & $b_1,c_1,b_2,c_2\neq 0 $ and all others equal to zero \\ 
			\hline
			$M_4$ & $b_1,c_1,b_2,c_2,b_3,c_3\neq 0 $ and all others equal to zero \\ 
			\hline
			$M_5$ & $b_1,c_1,b_2,c_2,b_3,c_3,b_4,c_4\neq 0 $ and all others equal to zero \\ 
			\hline
		\end{tabular} 
		\caption{Various models used in our analysis.}\label{tab:models} 
	\end{center}
\end{table}
\begin{table}
	\begin{center}
		
				\begin{tabular}{ |c|c|c|c| } 
			\hline
			\hline
			parameters & prior (uniform)&parameters & prior (uniform)\\
			\hline
			$\Omega_{dm}$ & [0.15,0.30] &$\alpha$& [0.10,0.16]\\
			\hline
			$\Omega_{ba}$ & [0.02,0.07]&$\beta$&  [2.85,3.25]\\
			\hline
			$h$ & [0.55,0.80]  & M & [-19.6,-19.2]\\
			\hline
			$\sigma_8$ & [0.50,1.20]  &$\Delta M$& [-0.15,0.15]\\
			\hline
			$b_1$ & [-0.3,0.3]  &$c_1$& [-0.3,0.3]\\
			\hline
			$b_2$ & [-0.2,0.2]  &$c_2$& [-0.2,0.2]\\
			\hline
			$b_3$ & [-0.1,0.1]  &$c_3$& [-0.1,0.1]\\
			\hline
			$b_4$ & [-0.05,0.05]  &$c_4$& [-0.05,0.05]\\
		\end{tabular}
		
		\begin{tabular}{ |c|c|c|c| } 
			\hline
			\hline
			parameters & prior (uniform)&parameters & prior (uniform)\\
			\hline
			$\Omega_{dm}$ & [0.0,0.60] &$\alpha$& [-0.3,0.3]\\
			\hline
			$\Omega_{ba}$ & [0.0,0.2]&$\beta$&  [1.0,4.0]\\
			\hline
			$h$ & [0.4,1.]  & M & [-21.,-18.]\\
			\hline
			$\sigma_8$ & [0.3,1.5]  &$\Delta M$& [-0.25,0.25]\\
			\hline
			$b_1$ & [-0.5,0.5]  &$c_1$& [-0.5,0.5]\\
			\hline
			$b_2$ & [-0.3,0.3]  &$c_2$& [-0.3,0.3]\\
			\hline
			$b_3$ & [-0.2,0.2]  &$c_3$& [-0.2,0.2]\\
			\hline
			$b_4$ & [-0.1,0.1]  &$c_4$& [-0.1,0.1]\\
			\hline
		\end{tabular}	
	\end{center}
	\caption{Two ranges of the model parameters which we consider in this work. The upper panel (lower panel) indicates a narrow(broad) prior.}\label{tab:prior} 
\end{table}  

First, the best-fit parameters and their uncertainties for all
of the models utilized in this analysis were obtained with the aid of
the MCMC method and the results are listed in Table (\ref{tab:res-mcmc}).
For comparison we also provide the results of 
the $\Lambda$CDM cosmological model, namely Pad\'{e} M$_{1}$ parametrization.
Notice that we use \emph{getdist} python package \footnote{https://github.com/cmbant/getdist} for the analysis of the MCMC samples. 
We find that the main cosmological 
parameters, namely $(\Omega_{m0},h,\sigma_{8})$ are practically the same for 
all models. Concerning the DE parameters it seems that 
although, the best fit values indicate $w<-1$ at the 
present time, we can not exclude the possibility of $w>-1$ at $1\sigma$ level. 
Moreover, for all Pad\'{e} parametrizations 
we find that the best fit values obey the inequalities
$b_{1}>c_{1}$ and $b_{2}>c_{2}$. 

Second we run the \emph{pymultinest} code in order to check 
the statistical performance of the current Pad\'{e} models 
in fitting the data and we compare them with that of $\Lambda$CDM. 
The minimum $\chi^2_{\rm min}$, Akaike Information Criterion (AIC), Bayesian Information Criterion (BIC) and the bayesian evidence (for two ranges of priors) of our 
models are summarized in Table (\ref{tab:res}). 
We remind the reader that the AIC \citep{Akaike:1974} and BIC \citep{Schwarz:1974} estimators 
are given by 
\begin{equation}
	{\rm AIC} = \chi^2_{\rm min} + 2n_{\rm fit},\\ 
	{\rm BIC} = \chi^2_{\rm min} + n_{\rm fit}\ln N, \;
\end{equation} 
where $n_{\rm fit}$ (N) is the number of fitted parameters (number of data points). 
Clearly, AIC identifies 
the statistical significance of our results, namely  
a smaller value of AIC implies a better model-data fit.
On the other hand, the model
pair difference $\Delta {\rm AIC}={\rm AIC}_{\rm model}-{\rm AIC}_{\rm min}$
provides 
the statistical performance of the different models in reproducing
the observational data.
Specifically, the condition
$4<\Delta {\rm AIC} <7$ indicate a positive evidence 
against  
the model with higher value of ${\rm AIC}$ \citep{Ann2002,Burnham2004}, while the 
inequality $\Delta {\rm AIC} \ge 10$ points a strong such evidence. 
Lastly, the restriction $\Delta {\rm AIC} \le 2$ provides an indication 
of consistency between the two comparison models.

Another way of testing the ability of the models to fit the 
data is via the Bayesian evidence ${\cal E}$, namely 
a model with the higher evidence is favored over another one. 
In this context, in order 
to measure the significant difference between two models $M_{i}$ and $M_{j}$
we can use the Jeffreys' scale \citep{jeff.book} which is given by 
$\Delta {\ln \cal E}={\ln \cal E}_{M_{i}}-{\ln \cal E}_{M_{j}}$. This 
model pair difference provides the following situations:
(i) $0<\Delta {\ln \cal E}<1.1$ suggests weak 
evidence against $M_{j}$ model when compared with $M_{i}$, (ii) the restriction 
$1.1<\Delta \ln {\cal E}<3$ means that there is definite
evidence against $M_{j}$, while   
in the case of $\Delta \ln {\cal E}\ge 3$ such evidence becomes strong \citep{Nesseris:2012cq}. 

In order to realize that extra parameters in a given model can be constrained by the data in hand, one can compute the so called Bayesian complexity which was introduced in \cite{Spiegelhalter:2002yvw}. The quantity measures the number of parameters that the data can support. Following \cite{Trotta:2008qt}, the Bayesian complexity given by
\begin{equation}\label{eq:Bay-com}
C_b \equiv \overline{\chi^2(\theta)}- \chi^2(\hat{\theta})
\end{equation}  
where bar indicates a mean taken over the posterior distribution and $\hat{\theta}$ can be either the best or the mean value of parameters . The Bayesian complexity is a quantity to measure the power of data to constrain parameters compare to the predictivity of the model which is given by the prior. Generally, the Bayesian complexity depends on both data and prior. But in our case, the sample of the free parameters are almost the same for two prior ranges, the Bayesian complexity does not depend on the prior.

Our estimation of Bayesian complexities are given in table (\ref{tab:res}). Here we consider both cases for $\hat{\theta}$, so $\hat{\theta_b}$ ($\hat{\theta_m}$) indicates the best value (mean value) of parameters. 

Clearly, after considering the above statistical tests
we find that the best model is the $\Lambda$CDM 
model, hence ${\rm AIC}_{\rm min}\equiv {\rm AIC}_{M_{1}}$,
${\cal E}_{M_{j}}\equiv {\cal E}_{M_{1}}$.
Using the model pair difference $\Delta {\rm AIC}$ we find 
strong evidence against models $M_{4}$ and $M_{5}$, namely
$\Delta {\rm AIC} \gtrsim 10$. Also, in the case of $M_{3}$ model
we have $\Delta {\rm AIC} \simeq 4.79$ which 
indicates positive evidence against that model, while 
for $M_{2}$ model we obtain $\Delta {\rm AIC} \simeq 1.13$ 
and thus we can not reject this model. 
In contrast to the AIC, BIC of our models indicate a "decisive" evidence against $M_2$,$M_3$ $M_4$ and $M_5$ models. The main reason is that, the BIC penalizes models with a high number of parameters more than AIC, specifically when there are large number of data points. 

From the viewpoint of $\Delta {\ln \cal E}$ \footnote{The relative uncertainties in log-evidence are of order of $0.1\%$ } we argue that there is a weak evidence in favor of all dynamical model  
 when compared with $M_{1}$ ($\Lambda$CDM). Note that we consider two different prior ranges to check the possible dependency of evidences on the prior (see table (\ref{tab:prior})). In fact, the evidence of each model is different considering different prior \footnote{The evidence for the $\Lambda$CDM with narrow prior is $-383.546\pm 0.24$ but with broad prior is $-393.782\pm 0.31$. } but $\Delta {\ln \cal E}$ (in our case) does not change significantly. In table (\ref{tab:res}) the results are presented for both narrow ($\Delta {\ln \cal E}_{N}$ ) and broad ($\Delta {\ln \cal E}_B$ ) priors.
 		
 		Of course such results are against Occam's razor, which simply penalizes models with a large number of free parameters. Models $M_5, M_4$ and $M_3$ have 8, 6 and 4 free parameters more than the $\Lambda$CDM but the Bayesian evidence does not show any significant difference between them. Similar to our conclusions can be found in the work of \citep{Nesseris:2012cq} 
in which they proved that a linear model $M_{a}$ with 14 free parameters
provides the same value of Bayesian evidence 
with another model $M_{b}$ which contains 4 free parameters.
According to these authors, 
the latter can be explained if the 
extra 10 parameters of $M_{b}$
do not really improve the statistical performance of 
the model in fitting the data.
In our case we confirm the results of \citep{Nesseris:2012cq} 
for Pad\'{e} cosmologies, namely the extra parameters of $M_3$, $M_4$ and $M_5$
parametrizations do not improve the corresponding DE models.

Moreover, the Bayesian complexity provides a diagnostic tool to break the degeneracy when two competing models have almost the same evidence. Since, from evidence alone, it is not clear that the extra parameters are unmeasured or improve the quality of the fit just enough to offset the Occam's razor penalty term, the Bayesian complexity can be used to break this degeneracy (for more information see \cite{Trotta:2008qt}). Our results indicate a slightly larger Bayesian complexity when two competing models are $M_1$ and $M_2$, which is consistent with our conclusion that current data slightly prefer a dynamical dark energy model. The Bayesian complexities for other models are more and less the same, which indicates current data are not good enough to measure  the additional parameters and extra parameters are not needed.

\begin{table*}
	\begin{center}
		\begin{tabular}{ |c|c|c|c|c|c| } 
			\hline
			\hline
			Models/Parameters & $M_1$&$M_2$ & $M_3$ & $M_4$ & $M_5$ \\
			\hline
		$\Omega_m$ &$0.2886\pm 0.0052$ &$0.2836^{+0.0052}_{-0.0059}$&$0.2830^{+0.0051}_{-0.0058}$&$0.2846^{+0.0048}_{-0.0054}$&$0.2843\pm 0.0056 $\\
			\hline
			$h$ &$0.6933\pm 0.0042$&$0.7026^{+0.0065}_{-0.0050}$& $0.7042^{+0.0066}_{-0.0059}$&$0.7023\pm 0.0058$& $0.7030\pm 0.0063$\\
			\hline
			$\sigma_8$ &$0.769\pm 0.023$&$0.766\pm 0.023 $&$0.765\pm 0.023$&$0.765\pm 0.023$& $0.765\pm 0.022$\\
			\hline
			$\alpha$ &$0.1413\pm 0.0050$&$0.1417\pm 0.0051$&$0.1417\pm 0.0050$&$0.1416\pm 0.0051$& $0.1417\pm 0.0051$\\
			\hline
			$\beta$ &$3.104\pm 0.057 $&$3.116\pm 0.058$&$3.117\pm 0.059$&$3.116\pm 0.058$& $3.114\pm 0.058$\\
			\hline
			$M$ &$-19.074\pm 0.016$&$-19.060\pm 0.018$&$-19.057\pm 0.018$&$-19.060\pm 0.017$& $-19.059\pm 0.017$\\
			\hline
			$\Delta M$ &$-0.071\pm 0.017 $&$-0.068^{+0.014}_{-0.018}$&$-0.067\pm 0.014$& $-0.067^{+0.014}_{-0.016}$& $-0.067^{+0.014}_{-0.016}$\\
			\hline
     		$b_1$ &-&$0.03^{+0.30}_{-0.17}$&$0.02^{+0.30}_{-0.27}$&$0.00\pm 0.23$& $0.01\pm 0.22$\\
			\hline
			$c_1$ &-&$-0.19^{+0.039}_{-0.093}$&$-0.14^{+0.13}_{-0.21}$& $-0.14^{+0.12}_{-0.22}$& $-0.14^{+0.11}_{-0.20}$\\
			\hline			
			$b_2$ &-&-&$0.03^{+0.18}_{-0.13}$& $0.01\pm 0.16$& $0.01\pm 0.16$\\
			\hline			
			$c_2$ &-&-&$-0.126^{+0.056}_{-0.16}$&$-0.095^{+0.067}_{-0.19}$& $-0.092^{+0.088}_{-0.18}$\\
			\hline
			$b_3$ &-&-&-&$0.039^{+0.15}_{-0.061}$& $0.017^{+0.14}_{-0.092}$\\
			\hline
			$c_3$ &-&-&-&$-0.04^{+0.19}_{-0.15}$& $-0.03\pm 0.11$\\
			\hline
			$b_4$ &-&-&-&-& $0.092^{+0.11}_{-0.075}$\\
			\hline
			$c_4$ &-&-&-&-& $0.005\pm 0.056$\\
			\hline															
		\end{tabular}
	\end{center}
	\caption{The best fit values and the corresponding 
$1\sigma$ uncertainties for the current Pad\'{e} parametrizations. Notice 
that the $\Lambda$CDM 
model can be seen as a Pad\'{e} $M_{1}$ parametrization, namely $b_{i}=c_{i}=0$.}\label{tab:res-mcmc} 
\end{table*}  
  
\begin{table*}
	\begin{center}
		\begin{tabular}{ |c|c|c|c|c|c|c|c| } 
			\hline
			\hline 
		    Model & $\chi^2_{\rm min}$& $\Delta$AIC& $\Delta$BIC & $\Delta {\rm \ln} {\cal E}_N $&$\Delta {\rm \ln} {\cal E}_B $& $C_b(\hat{\theta}_b)$&$C_b(\hat{\theta}_m)$\\
			\hline
			$M_1$ & 733.20 &0.0&0.0& 0.00&0.00&9.09&8.89\\
			\hline
			$M_2$ &730.33&1.13& 10.51 &0.22&0.55&9.92&9.33\\
			\hline
			$M_3$ &729.99 &4.79& 23.55 &0.65&0.82&9.85&9.62\\
			\hline
			$M_4$ &729.95&8.75& 36.89&0.97&0.68&9.71&9.32\\
			\hline
			$M_5$ &729.85&12.65& 50.17&1.02&0.95&9.77&9.6\\
			\hline
		\end{tabular}
	\end{center}
	\caption{The goodness-of-fit statistics 
$\chi^{2}_{min}$, $\Delta$AIC, $\Delta$BIC, $\Delta{\rm ln}{\cal E}$ for the narrow and broad ranges of priors ($N$ ($B$) stands for narrow(broad) prior ranges) and the Bayesian complexity for our models.}\label{tab:res} 
\end{table*}


Combining the aforementioned results we argue 
that although, the $\Lambda$CDM model reproduces 
very well the cosmological data, 
the possibility of a dynamical DE 
in the form of $M_{2}$ Pad\'{e} model 
cannot be excluded by the data.

\section{Conclusion}\label{sec:con}
In this article we attempt to check whether 
a dynamical dark energy is allowed by 
the current cosmological data. The evolution of dark energy 
is treated within the context of Pad\'{e} parameterization which can be seen 
as an expansion around the usual $\Lambda$CDM cosmology.
Unlike most DE parameterizations (CPL and the like), in the case of 
Pad\'{e} parametrization the equation of state parameter
does not diverge in the far future ($a\gg 1$)
and thus its evolution is smooth in the range of $a\in (0,+\infty)$.

Using the latest cosmological data 
we placed observational
constraints on the viable Pad\'{e} dark energy models, by 
implementing a joint statistical analysis 
involving the latest observational data, 
SNIa (JLA), BAOs, direct measurements of $H(z)$, 
CMB shift parameters from Planck and growth rate data.
In particular, we considered four Pad\'{e} parametrizations, each with several
independent parameters and we found that
practically, the examined Pad\'{e} models, are
in very good agreement with observations.
In all of them the main cosmological 
parameters, namely $(\Omega_{m0},h,\sigma_{8})$ are practically the same. 
Regarding the free parameters of Pad\'{e} parametrtization we showed 
that although, the best fit values indicate $w<-1$ at the 
present time, we can not exclude the possibility of $w>-1$ at $1\sigma$ level. 

Finally, for all Pad\'{e} models 
we quantified their deviation from $\Lambda$CDM cosmology
through AIC and Jeffreys' scale. 
We found that the corresponding $\chi^{2}_{\rm min}$ values 
are very close to that of 
$\Lambda$CDM which implies that the chi-square estimator can not distinguish 
Pad\'{e} models from $\Lambda$CDM.
Among the family of the current 
Pad\'{e} parametrizations, the model which contains two dark energy 
parameters is the one 
for which a small but non-zero deviation from
$\Lambda$CDM cosmology is slightly allowed by AIC test.
On the other hand, based on Jeffreys' scale  
we showed that a deviation from $\Lambda$CDM cosmology
is also allowed, hence 
the possibility of a dynamical DE 
in the form of Pad\'{e} parametrization 
cannot be excluded.
 
 Furthermore, we estimated the so called Bayesian complexity to realize whether current data can constrain the extra parameters in our models or not. The Bayesian complexity provides a measurement of the effective number of parameters, which can be measured, given the data, and our results showed a slightly lager Bayesian complexity for $M_2$ model which is consistent with our conclusion about possible dynamical DE models. In contrast, the Bayesian complexity does not change significantly by adding extra parameters in our other models which indicates that current data is not good enough to measure the extra parameters.       

 \bibliographystyle{spphys}
\bibliography{ref}

\label{lastpage}

\end{document}